\documentclass[12pt]{iopart}
\usepackage{graphicx}

\usepackage{color}

\newcommand\acie{Angew. Chem. Int. Ed.}
\newcommand\arbb{Ann. Rev. Biophys. Bioeng.}
\newcommand\bc{Biochemistry}
\newcommand\cbc{ChemBioChem}
\newcommand\cpc{Comp. Phys. Comm.}
\newcommand\jacs{J. Am. Chem. Soc.}
\newcommand\jasa{J. Am. Stat. Assoc.}
\newcommand\jasms{J. Am. Soc. Mass Spectrom.}
\newcommand\jcp{J. Chem. Phys.}
\newcommand\jpca{J. Phys. Chem. A}
\newcommand\jpcb{J. Phys. Chem. B}
\newcommand\jpcl{J. Phys. Chem. Lett.}
\newcommand\molp{Mol. Phys.}
\newcommand\pmag{Philos. Mag.}
\newcommand\pmsix{Philos. Mag. Ser. VI}
\newcommand\pnasu{Proc. Natl. Acad. Sci. USA}
\newcommand\prc{Phys. Rev. C}
\newcommand\prb{Phys. Rev. B}
\newcommand\prl{Phys. Rev. Lett.}

\begin{document}

\title{Structure and stability of charged clusters}

\author{Mark A.~Miller}
\address{University Chemical Laboratory, Lensfield Road,
Cambridge CB2 1EW, United Kingdom}
\ead{mam1000@cam.ac.uk}

\author{David A.~Bonhommeau}
\address{GSMA UMR7331, Universit{\'e} de Reims Champagne-Ardenne,
UFR Sciences Exactes et Naturelles, Moulin de la Housse B.P.~1039,
51687 Reims Cedex 2, France}

\author{Christopher J.~Heard}
\address{School of Chemistry, University of Birmingham,
Edgbaston, Birmingham B15 2TT, United Kingdom}

\author{Yuyoung Shin}
\address{University Chemical Laboratory, Lensfield Road,
Cambridge CB2 1EW, United Kingdom}

\author{Riccardo Spezia}
\address{Universit{\'e} d'Evry val d'Essonne, LAMBE CNRS UMR8587,
Blvd F.~Mitterrand, B{\^a}t Maupertuis, 91025 Evry, France}

\author{Marie-Pierre Gaigeot}
\address{Universit{\'e} d'Evry val d'Essonne, LAMBE CNRS UMR8587,
Blvd F.~Mitterrand, B{\^a}t Maupertuis, 91025 Evry, France}
\address{Institut Universitaire de France (IUF), 103 Blvd St Michel,
75005 Paris, France}
\ead{mgaigeot@univ-evry.fr}

\date{\today}

\begin{abstract}
When a cluster or nanodroplet bears charge, its structure and thermodynamics are
altered and, if the charge exceeds a certain limit, the system becomes
unstable with respect to fragmentation.  Some of the key results in this
area were derived by Rayleigh in the nineteenth century using a continuum
model of liquid droplets.
Here we revisit the topic using a simple particle-based description,
presenting a systematic case study of how charge affects the physical
properties of a Lennard-Jones cluster composed of 309 particles.  We find that
the ability of the cluster to sustain charge depends on the number of particles
over which the charge is distributed---a parameter not included in Rayleigh's
analysis.  Furthermore, the cluster may fragment before the charge is strong
enough to drive all charged particles to the surface.
The charged particles in stable clusters are therefore likely to reside
in the cluster's interior even without considering solvation effects.
\end{abstract}

\maketitle

\section{Introduction}
Charged clusters and small droplets are important in many areas of
physical chemistry, ranging from aerosol formation in the atmosphere \cite{Hirsikko11a}
to electrospray processes in the laboratory \cite{Fenn03a}.  The presence of charge
on a droplet affects both its interactions with other species and the physical
properties of the individual droplet.  Amongst the most important questions relating
to physical properties are how much charge a cluster or droplet can accommodate
without becoming unstable on short timescales, and how the break-up of the droplet
proceeds as this charge limit is approached and exceeded.
\par
One area where such
considerations are crucial is electrospray ionisation mass spectrometry.  The
electrospray technique produces a stream of charged droplets that
solvate molecules or complexes
to be conveyed into the mass spectrometer for analysis.  Before the analyte's
mass-to-charge ratio is measured, it becomes separated from the solvent, and some of
the droplet's charge is deposited on it.
Various models have been proposed to describe the mechanism by which the solvent leaves the
analyte \cite{Dole68a,Iribarne76a,Konermann09a} in different circumstances.
Computer simulations with atomistic detail have been employed to obtain a more
explicit picture of the desolvation process and also to study the structure of
water droplets containing ions \cite{Ahadi09a,Ahadi10a,Ichiki06a}
and solvated biomolecules \cite{Consta10a,Marginean06a,Steinberg08a,Patriksson07a}.
\par
On the fundamental question of the stability of pure charged droplets, the key theoretical
work dates back well over a hundred years to Rayleigh \cite{Rayleigh82a}, who considered
arbitrary deformations of a uniformly charged, structureless spherical droplet.  Deformations
away from a spherical shape increase the surface area of the droplet, thereby increasing
its surface energy, but at the same time decrease the electrostatic repulsion by spreading
the charge out.  At a total charge
\begin{equation}
Q_{\rm R} = \pi\sqrt{8\epsilon_0\gamma D^3},
\label{Rayleigh}
\end{equation}
known as the Rayleigh limit, the electrostatic contribution
to the energy overwhelms the surface energy and the droplet becomes mechanically unstable.
In Eq.~(\ref{Rayleigh}), $\gamma$ is the surface tension of the liquid, $\epsilon_0$ is
the vacuum permittivity and $D$ is the droplet diameter.  Rayleigh's formula remains
of great practical use to the present day.
\par
To examine the nature of charge-driven instabilities with particle-level detail,
simulations of small charged clusters have been carried out by adding Coulomb interactions
to simple generic potentials like Lennard-Jones \cite{Ison04a} and Morse \cite{Levy06a}.
However, these studies have mostly concentrated on total charges where the cluster is
decisively in the unstable regime and rapidly undergoes fragmentation.
\par
In this article, we take a closer look at the properties of charged clusters with
an explicit but generic representation of individual particles,
concentrating on the structure of clusters as the
charge-driven instability is approached from below.  We treat the clusters at equilibrium,
setting aside the dynamical aspects of fragmentation for now.  The results presented here
constitute a case study on a cluster of 309 Lennard-Jones particles to which Coulomb repulsion
has been added.  309 is a ``magic number'' for the Lennard-Jones potential; the lowest-energy
structure consists of four complete icosahedral shells around a central atom \cite{Romero99a}.
The present study represents the preliminary steps in
a more comprehensive ongoing investigation of charged droplets.

\section{Model and methods}
Our cluster of $N=309$ particles is bound by the pairwise potential
\begin{equation}
V=4u\sum_{i<j}^N\left[\left(\frac{\sigma}{r_{ij}}\right)^{12}
-\left(\frac{\sigma}{r_{ij}}\right)^6\right]+
\sum_{i<j}^N\frac{q_i q_j}{4\pi\epsilon_0 r_{ij}},
\label{potential}
\end{equation}
where $u$ and $\sigma$ are the Lennard-Jones pair well-depth and diameter, respectively,
$r_{ij}$ is the separation of particles $i$ and $j$, and $q_i$ is the charge on particle $i$.
By defining a reduced charge $q_i^*=q_i/(4\pi\epsilon_0\sigma u)^{1/2}$,
along with the usual Lennard-Jones reduced energy and lengths $V^*=V/u$ and
$r_{ij}^*=r_{ij}/\sigma$,
Eq.~(\ref{potential}) may be rewritten in the dimensionless form
\begin{displaymath}
V^*=\sum_{i<j}^N\left[4({r_{ij}^*}^{-12}-{r_{ij}^*}^{-6})+q_i^* q_j^*/r_{ij}^*\right].
\end{displaymath}
We will consider clusters where $n$ of the particles carry a charge $q_i^*=q^*$ and the remaining
$N-n$ particles are uncharged, making the overall charge on the cluster $Q^*=nq^*$.
Although $q$ itself should be an integer multiple of the electronic charge $e$, the
reduced charge $q^*$ also depends on the Lennard-Jones parameters.  We will therefore
treat $q^*$ as a continuously variable effective charge.
The fact that the particles constituting the cluster are represented explicitly
and may each carry a charge of only $q^*$ or $0$
means that there is no need to consider an underlying continuum
medium with its own conductivity.  The charge distribution is completely determined by
the function $V^*$ and a small number of external parameters, such as the temperature.
Therefore, although our model is coarse-grained, its particulate nature makes it more physically
detailed than a continuum model.
\par
We simulate the cluster at thermodynamic equilibrium
using canonical Monte Carlo (MC) simulations.  In addition to standard
displacements of individual particles, we employ trial moves that attempt to swap the charges
on randomly chosen charged--uncharged pairs of particles while keeping all positions fixed.
These steps are accepted according to the usual Metropolis criterion \cite{Frenkel02b}.
The swap moves are not intended to mimic the dynamical process of charge transfer
or conductivity that
may occur between molecules in certain systems.  In particular, the swaps are not restricted
to pairs of adjacent particles.  The purpose of the moves is to enhance the efficiency
with which the equilibrium distribution of charges is reached and sampled, and the moves achieve
this by, in effect, decoupling the mobility of the charge from the movement of individual
particles.  The same distribution would eventually be achieved by any set of MC
moves that obey (detailed) balance.
\par
We will be concerned with the behaviour of the cluster as a function of the magnitude of the
single-particle charge $q^*$ for a given $n$ and, to some extent, on the properties with
a given charge as a function of temperature $T$.  To accelerate sampling in these cases,
we have employed replica-exchange MC \cite{Geyer95a}, using either the temperature \cite{Neirotti00a}
or the charge as the parameter that varies between replicas.  In the latter case, the
acceptance criterion for exchanging configurations ${\bf X}_1$ and ${\bf X}_2$ that
are initially found in
replicas where the single-particle charges are $q^*_a$ and $q^*_b$, respectively, is
\begin{displaymath}
p^{\rm acc} = {\rm min}\left\{1,\exp\left[\left(W({\bf X}_1)-W({\bf X}_2)\right)
({q_a^*}^2-{q_b^*}^2)/kT\right]\right\}.
\end{displaymath}
Here, $k$ is Boltzmann's constant and $W({\bf X})=\sum_{i<j}^n {r_{ij}^*}^{-1}$,
where the sum runs over pairs of charged particles.
\par
Any cluster is thermodynamically unstable with respect to vaporisation unless it is
confined.  We therefore constrain the cluster to lie within a spherical container
of radius $R=6.5\sigma$ whose centre follows the centre-of-mass of the
cluster \cite{Lee73a}.  This radius is large enough not to inhibit large fluctuations
in the shape of the cluster and also allows fragmentation to occur.  We shall
return to the consequences of this particular choice of radius when discussing the results.

\section{Stability}
\begin{figure}
\centerline{\includegraphics[width=75mm]{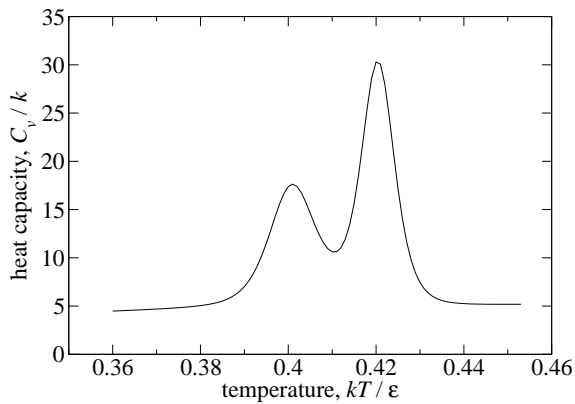}}
\caption{Heat capacity of the neutral LJ$_{309}$ cluster, obtained by the multiple histogram
method.
\label{cv}}
\end{figure}

We will consider the effect of adding charge to the Lennard-Jones cluster at reduced temperatures
$T^*=kT/u=0.2$ and $0.43$.  Figure \ref{cv} shows the canonical
constant-volume heat capacity $C_v$ of the cluster, derived from temperature replica-exchange
simulations coupled with multiple histogram reweighting \cite{Labastie90a,Weerasinghe93a}.
The taller of the two peaks is due to the main melting transition \cite{Noya06a},
meaning that our two temperatures correspond to well within the solid-like state and just within the
liquid-like regime, respectively.

\begin{figure}
\centerline{\includegraphics[width=75mm]{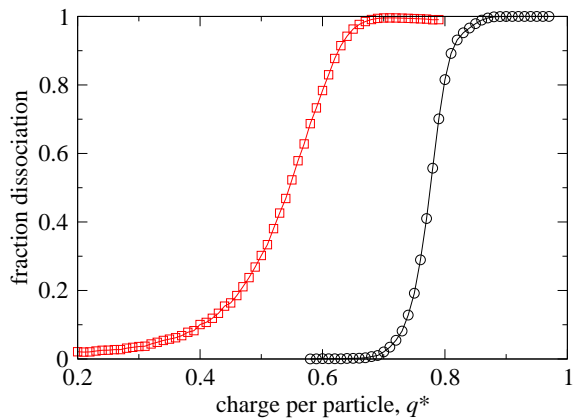}}
\caption{(Colour online)  Fraction of configurations in which dissociation is detected
for LJ$_{309}$ with $n=100$ charged particles at temperatures $T^*=0.2$ (black circles) and
$0.43$ (red squares) as a function of the charge per particle.
\label{dissoc}
}
\end{figure}

At $T^*=0.2$, the neutral cluster fluctuates uneventfully about its icosahedral global
minimum \cite{Romero99a}.  This qualitative behaviour persists if some charge is added,
but for sufficiently large charge, a small number of charged particles are expelled,
leaving a well-defined sub-cluster at the centre of the spherical container.  The
expelled particles typically lie close to the container edge, repelled there by the
long range Coulomb potential.  We may therefore count an individual configuration
as dissociated if at least one
charged particle lies within a small distance, chosen to be $\sigma/2$,
from the container wall.  The line marked with circles in Fig.~\ref{dissoc} shows that
the equilibrium
fraction of dissociated configurations at equilibrium rises sharply at a particular
value of the single-particle charge, indicating that electrostatic repulsion has
become strong enough to overcome the cohesive part of the Lennard-Jones potential.
We will designate the total charge at which
the fraction of dissociated configurations passes through $0.5$ by $Q_{\rm max}$; beyond
this threshold, more than half the equilibrium configurations are dissociated.

At the liquid-like temperature $T^*=0.43$, a similar pattern is observed, as shown by
the squares in Fig.~\ref{dissoc}.  However, since the cluster has now escaped the basin
of attraction of its icosahedral global minimum and explores a range of
disordered structures, the transition from an intact cluster to
full dissociation is a broader function of the charge.  We will nevertheless use the same
criterion of $Q_{\rm max}$ to quantify the maximum charge that the cluster can sustain.
\par
\begin{figure}
\centerline{\includegraphics[width=75mm]{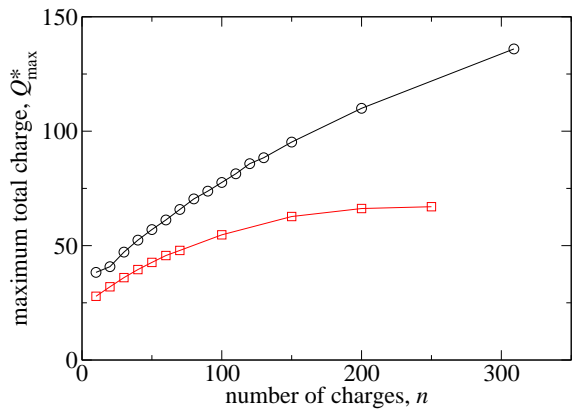}}
\caption{(Colour online)  Maximum total charge $Q^*_{\rm max}$ that the cluster can
sustain at $T^*=0.2$ (black circles) and $T^*=0.43$ (red squares) as a function of the number
of particles $n$ over which the charge is spread.
\label{maxcharge}
}
\end{figure}

Importantly, $Q_{\rm max}$ is a function of $n$, {\it i.e.}~the maximum total charge that the
cluster can sustain depends on the number of particles over which the charge is spread.
Figure \ref{maxcharge} shows that $Q_{\rm max}$ increases quite dramatically with $n$,
especially at the lower temperature.  It is not straightforward to compare
the $Q_{\rm R}$ of Eq.~(\ref{Rayleigh}) with $Q_{\rm max}$, not least because
we must work at temperatures well below the triple point of
$T^*=0.694$ \cite{Mastny07a}, where the bulk surface tension is not known.
However, we note that the influence
of $n$ discrete charges does not enter into Rayleigh's analysis because in that analysis
the droplet is treated as being homogeneously charged.  The dependence of the maximum
charge on $n$ is
potentially important in the context of small electrospray droplets, where the charge
of the droplet may be due to a relatively small number of ions.  Our thermodynamically
based $Q_{\rm max}$ suggests that such droplets have a greater tendency to dissociate
by the emission of charged particles than a hypothetical cluster where the same amount
of charge is spread out over a larger number of particles.

Increasing the container radius $R$ decreases the value of $Q_{\rm max}$ that is obtained
because dissociated charged particles can then go further from the remaining sub-cluster,
thereby lowering the potential energy of the configuration and raising its statistical
weight relative to the undissociated cluster.  However, the decisive trend of increasing
$Q_{\rm max}$ with $n$ persists for all $R$ large enough to permit dissociation.
\par
If the charge is increased beyond $Q_{\rm max}$, a further structural change may result.
It is possible for the cluster to split into two or more large sections, which then
repel each other as far as the container will allow.  This subdivision does not happen
if $n$ is sufficiently small since the electrostatic energy can then be lowered by the
emission of individual particles without sacrificing the attractive Lennard-Jones
interactions in the main part of the cluster.  However, as $n$ increases, subdivision
occurs at a threshold increasingly close to $Q_{\rm max}$ and, for large container radius,
may even preempt significant loss of individual charged particles.
This observation again points to the importance of taking into account not only the total
charge borne by a cluster but also the number of particles over which
the charge is divided.  One may anticipate that a dynamic study would find increased
competition between different instability mechanisms at higher $n$.

\section{Structure}

An important question concerning charged droplets is where the charged particles reside.
In the case of ions dissolved in water, highly specific effects come into play
in determining the affinity of ions for the surface of the bulk liquid itself,
according to how a given ion influences the interactions between
the water molecules \cite{Caleman11a,Takahashi11a}.  In a droplet and
in the absence of strong
surface affinity effects, one might expect ions to be mutually repelled to
the droplet surface by electrostatic considerations.  However, this is not
always found to be the case \cite{Znamenskiy03a,Ahadi10a}.
\par
Our model is more straightforward and generic than water, allowing us to consider 
the structure of charged droplets without the rather special effects of hydrogen
bonding.  Energetically, the driving forces in our model are towards the
optimisation of as many Lennard-Jones pairwise neighbours as possible and
toward the maximisation of distance between like charges.  The global potential
energy minimum structure of the cluster, obtained using the basin hopping
algorithm \cite{Wales97a} is always
a slightly relaxed version of the neutral icosahedron \cite{Romero99a}.  As $n$
is increased from small values, the charged particles preferentially occupy
vertices, then edges
and finally faces of the outer shell before populating the cluster's interior.
Counteracting these energetic tendencies at finite temperature is the entropy
of permuting charged and neutral particles.
\par
To determine how the charges are divided between the surface and the interior of
the cluster, we need a method for identifying surface particles.  At solid-like
temperatures, where the cluster retains a clear structure of concentric shells,
the outermost shell of 162 particles can unambiguously be designated as the surface.
For liquid-like structures, the surface is less clear-cut and its curvature adds
to the difficulty of devising a simple algorithm to detect the outermost layer.
Existing methods to tackle this problem include the cone algorithm of Dellago
and coworkers \cite{Wang05b}, which identifies surface particles as those for
which the vertex of a cone with a particular aperture can be placed at the particle's
centre without any other particles falling inside the cone.  The cone may have
any orientation, thereby taking into account the fact that the surface normal
of a droplet may locally deviate significantly from the vector joining that region
to the droplet's centre.
\par
We have devised a somewhat more straightforward algorithm that shares the advantageous
features of the cone algorithm and produces a reliable and intuitively realistic
division between surface and core particles.  The method borrows an idea from algorithms
used to determine the surface of proteins \cite{Richards77a}.  The principle is that a particle
should be regarded as lying on the surface if it would be accessible by a fictitious
probe particle approaching from a distance, while the cluster configuration is held fixed.
To determine this accessibility,
a nominal hard-sphere diameter $D_{\rm LJ}$ is assigned to the Lennard-Jones particles.
We then attempt to place a probe particle of diameter $D_{\rm pr}$ in contact with
any combination of three particles that are sufficiently close for three contacts to
be achieved simultaneously.  For each such triplet, there are two possible positions
for the probe particle, located symmetrically on either side of the plane defined by
the triplet on the line passing through the circumcentre of the triangle.  If the probe
particle in at least one of these positions does not overlap with any other particle in
the cluster then all three Lennard-Jones particles are designated as lying on the surface.
A given particle may qualify as being on the surface by being tested in conjunction with
any two of its neighbours.  Although a systematic test over all triplets of particles
may seem computationally expensive, it is quite easy to omit needless iterations of the
outer loops, and neighbour lists \cite{Allen87a} could be implemented for efficiency in
larger systems if necessary.
\par
\begin{figure}
\centerline{\includegraphics[width=60mm]{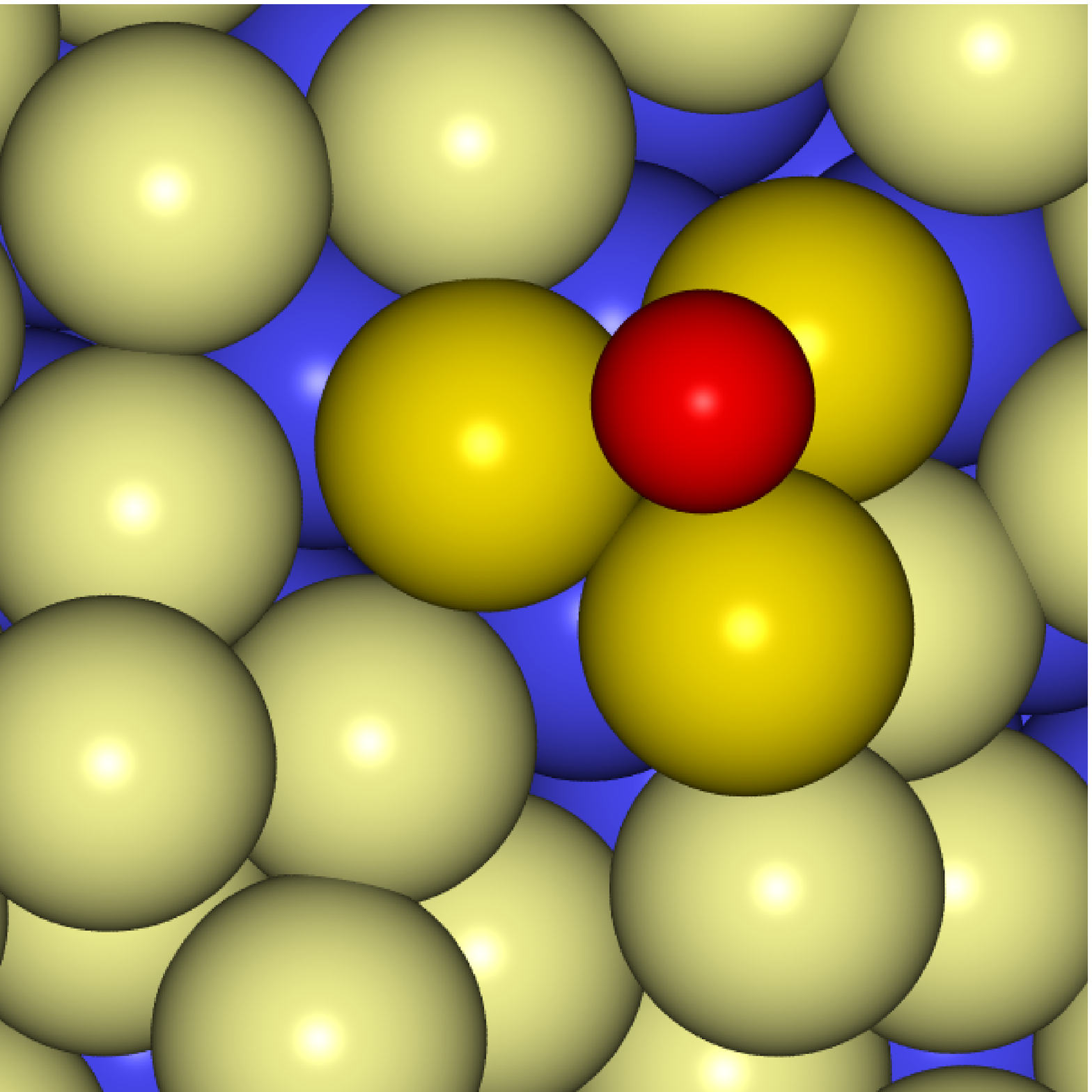}}
\centerline{\includegraphics[width=60mm]{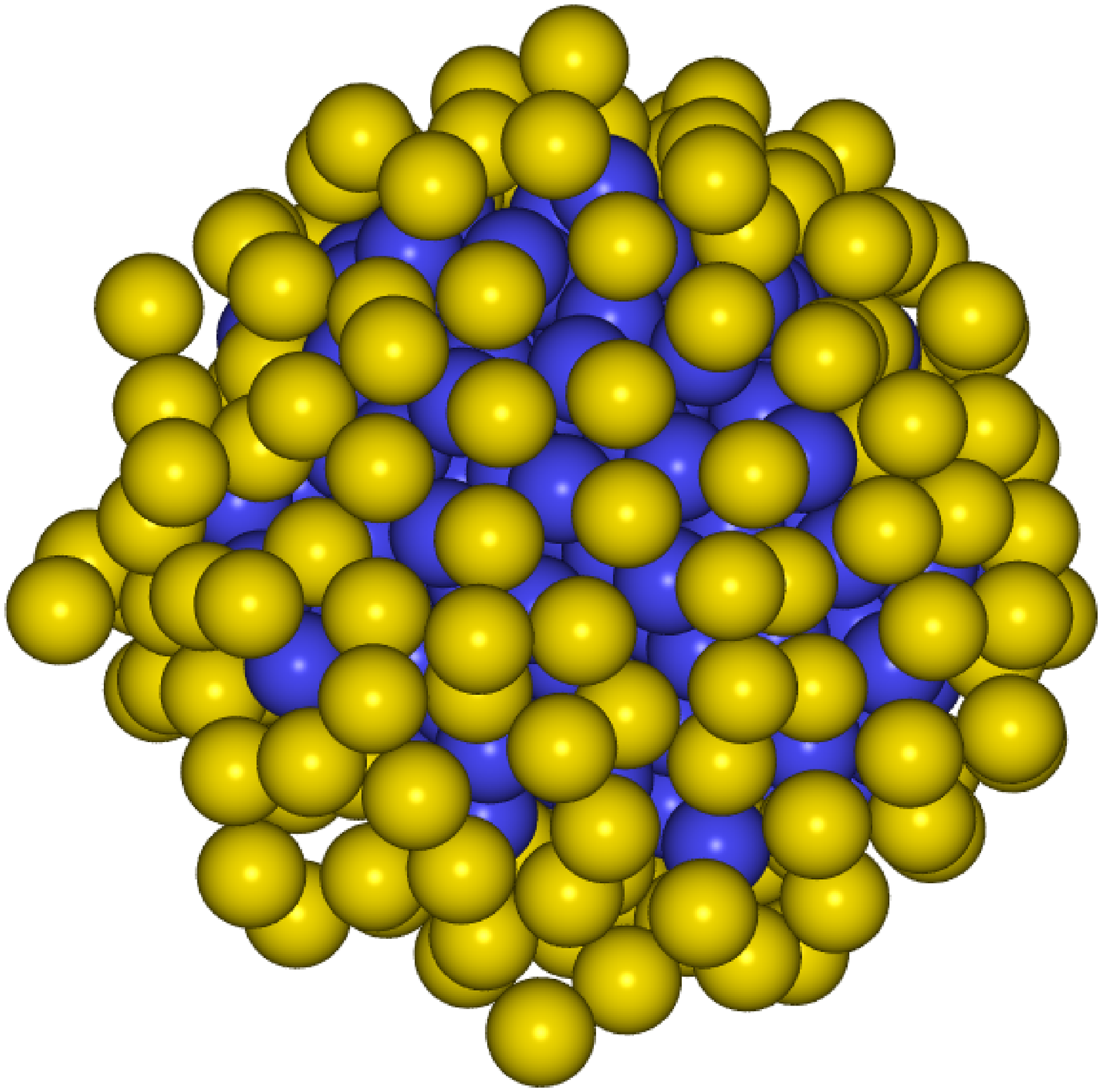}}
\caption{(Colour online) Top: Illustration of the method for distinguishing surface
particles (light/yellow) from the core of the cluster (dark/blue).  The surface particles
touching the probe (small, red) are highlighted.  Bottom: liquid-like
snapshot of the cluster, highlighting the surface particles.
\label{surface}
}
\end{figure}

The only parameters to fix in this probe method are the nominal diameters $D_{\rm LJ}$
and $D_{\rm pr}$.  Both should be comparable with $\sigma$ and the results should not be
sensitive to the values chosen.  We have found that choosing a slightly smaller probe
diameter enables the algorithm to cope better with crevices in the surface and adopted
$D_{\rm LJ}=1.2\sigma$, $D_{\rm pr}=0.8\sigma$ in this study.
The probe method and an example of its outcome are illustrated
in Fig.~\ref{surface}.  

\begin{figure}
\centerline{\includegraphics[width=75mm]{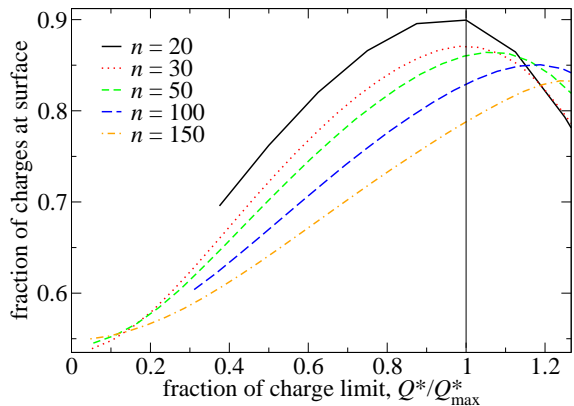}}
\caption{(Colour online) The fraction of charges that occupy surface sites as a function
of the total charge of the cluster for different numbers $n$ of charged particles at
temperature $T^*=0.43$.  For each value of $n$, the charge axis has been scaled by the
maximum charge from Fig.~\ref{maxcharge}.  The vertical line highlights the limit
$Q^*=Q^*_{\rm max}$.
\label{occupancy}
}
\end{figure}

Figure \ref{occupancy} shows how the average number of charged particles occupying surface sites
varies with the magnitude of the charge for several values of $n$ at $T^*=0.43$.  Due to
the liquid-like structure, the number of surface sites fluctuates and the sites have
to be identified for each sample separately.  The
charge on the horizontal axis of the plot has been scaled by the maximum charge that
the cluster can sustain for each value of $n$ respectively (Fig.~\ref{maxcharge}).
Looking at the vertical
line $Q^*/Q^*_{\rm max}$ shows that it is only for very small numbers of charges and only
when the charge approaches the limit of stability of the droplet
that one finds most of the charged particles on the surface of the cluster.
Hence, for a cluster of this size bound by the generic Lennard-Jones potential, the
charge required to drive the majority of particles electrostatically
to the cluster surface is already too much for the cluster itself to remain intact.
Therefore, even in a model without specific surface affinity effects or polarisability, charged
particles may be found predominantly in the interior of the cluster.

For $Q^*>Q^*_{\rm max}$, the loss of charged particles from the cluster causes the
number of charges at the surface to start decreasing for small $n$ in Fig.~\ref{occupancy}.
However, for higher $n$, although some charged particles are expelled, the higher charge
still leads to an overall increase in the proportion of charges driven to the surface of
the remaining sub-cluster, and the upward trend continues beyond $Q_{\rm max}$.
\par
In our simple model, the occupation of surface sites by charged particles is determined
by a competition between electrostatic repulsion of charges to the surface and the entropy
of mixing of charged and uncharged particles.  For solid-like temperatures, the balance
between these effects can be calculated quite accurately with little effort by using a few
approximations.  At these low temperatures, let us neglect the effect of thermal
fluctuations on the positions of the particles and consider the structure to be the
perfect icosahedron of the neutral cluster's global potential energy minimum, which has
$S=162$ surface particles.  If $n$ of the $N=309$ particles are charged, the number of
ways of permuting $s$ of them amongst the surface sites and the remaining $n-s$ amongst
the $N-S$ core sites is
\begin{displaymath}
\Omega_{N,n}(s) = C^S_s\, C^{(N-S)}_{(n-s)},
\end{displaymath}
where $C^a_b=a!/[b!(a-b)!]$ is a binomial coefficient.  Let us also define
\begin{equation}
W_{N,n}(s)=\left\langle\sum_{i<j}^n {r_{ij}^*}^{-1}\right\rangle_s,
\label{average}
\end{equation}
where the sum runs over the charged particles only and the angle brackets imply an
unweighted average over configurations in which $s$ charges are
randomly assigned to surface sites in
the frozen cluster and the remaining charges are randomly distributed amongst the
core sites.  $W_{N,n}(s)$ can be sampled very quickly since the particle positions
are held fixed while the charges are permuted amongst the sites,
and the reciprocal pairwise distances need be calculated only once.
The mean energy of configurations with $s$ charges at surface sites is then given
instantly for any magnitude of the charge by
\begin{displaymath}
{\bar V}^*_{N,n}(s) = {q^*}^2 W_{N,n}(s).
\end{displaymath}
We may therefore predict the probability that precisely $s$ particles lie on surface
sites to be
\begin{equation}
p_{N,n}(s) \propto \Omega_{N,n}(s)\,\exp\left[-{\bar V}^*_{N,n}(s)/kT\right],
\label{surfprob}
\end{equation}
with the constant of proportionality defined by the normalisation
$\sum_{s=0}^n p_{N,n}(s)=1$.
\par
\begin{figure}
\centerline{\includegraphics[width=75mm]{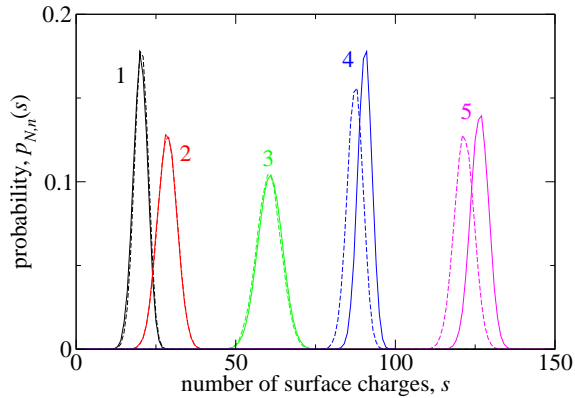}}
\caption{
Probability distributions of the number of charged particles lying on the cluster
surface for various combinations of parameters as listed in the legend.
For each pair of curves, the solid line is from explicit MC calculations while the
dashed line is the prediction of Eq.~(\ref{surfprob}).  The five sets of data
correspond to the following parameters:\\
1: $n=30$, $q^*=0.33$, $T^*=0.2$, $Q^*/Q^*_{\rm max}=0.28$\\
2: $n=50$, $q^*=0.16$, $T^*=0.3$, $Q^*/Q^*_{\rm max}=0.20$\\
3: $n=100$, $q^*=0.14$, $T^*=0.2$, $Q^*/Q^*_{\rm max}=0.26$\\
4: $n=100$, $q^*=0.44$, $T^*=0.2$, $Q^*/Q^*_{\rm max}=0.80$\\
5: $n=150$, $q^*=0.336$, $T^*=0.2$, $Q^*/Q^*_{\rm max}=0.80$
\label{prediction}
}
\end{figure}

Figure \ref{prediction} compares the predictions of Eq.~(\ref{surfprob}) with the
results of explicit MC simulations.
The averages for $W_{N,n}(s)$ in Eq.~(\ref{average})
were made using $10^5$ samples at each value of $s$, which
allows $W_{N,n}(s)$ to be evaluated in a matter of seconds for the full range of $s$.
At low $n$ and modest $Q^*/Q^*_{\rm max}$,
the predictions of Eq.~(\ref{surfprob}) compare essentially perfectly
with the simulations even at $T^*=0.3$, which is approaching the first
peak in the heat capacity (Fig.~\ref{cv}).  As $n$ increases, Eq.~(\ref{surfprob})
begins to overestimate the number of charged particles at the surface, but the
peak position is only in error by a few per cent, even for $n=150$ and
a total charge approaching $Q_{\rm max}$.  Given that
${\bar V}^*_{N,n}$ is a crude average and that thermal motion and charge-induced
distortion of the cluster are neglected in
Eq.~(\ref{surfprob}), the simple and fast prediction works well.  Importantly,
we may conclude that the opposing driving forces of electrostatic repulsion and
the permutational entropy capture most of the effects that determine the location
of the charges in this model at low temperatures.
\par
In principle, a similar approach could be taken to the liquid-like state.
However, the number of surface sites is then no longer fixed and it may be necessary
to sample a selection of disordered structures.  Furthermore, we may anticipate that
the presence of charge will have a greater influence in distorting the cluster from
its neutral shape than it does in the solid-like regime.  We are currently investigating
the influence of these effects.
\par
We conclude this section by noting that in some systems there may be more sophisticated
mechanisms that affect the equilibrium distribution of charges.  For example, the surface
of liquid water may develop a slight excess charge even without dissolved ions due to the
asymmetry of hydrogen bonds, which results in a small transfer of charge between molecules
\cite{Vacha11a,Vacha11b}.
A related effect can lead to some of the charge of a solvated ion appearing near the
water surface via a chain of partial charge transfers through oriented water
molecules \cite{Ahadi10a}.  Hence, some of the {\em charge} may appear at the surface
even though the {\em ions} are located within the bulk liquid.
Our simple Lennard-Jones model does not attempt to capture the
structure of water or the rather specific effects associated with hydrogen bonding.
However, the model could be modified to incorporate some degree of charge
separation within individual particles by adding a polarisability term to the
potential.  Self-consistent solution of the polarisability equations to obtain the
induced dipole moments adds a considerable computational overhead to
the simulations.  Some preliminary calculations of this sort reveal, as expected, that
polarisability leads to an even lower tendency of charged particles to lie at the surface
of the droplet.

\section{Concluding remarks}

We have taken a thermodynamic approach to the stability and structure of the
309-particle Lennard-Jones cluster bearing charge on some of the particles.  By confining
the cluster to a spherical container we have calculated the equilibrium between intact
and dissociated configurations and have obtained statistics for the occupation of surface
sites by charged particles.
\par
The key observation from this work is that both the stability of the cluster and
the extent to which charges are likely to be found at its surface depend not just on the
total charge but also on the number of particles over which the charge is distributed.
When the charge is spread over more particles, the cluster is able to sustain a greater
total charge but the fraction of the charges that reside at surface sites is lower.
In general, the cluster is prone to break up at charges lower than those required to
drive all the charged particles to the surface.
\par
The location of charged particles in a sphere became an important topic following the
introduction of the ``plum pudding'' model of the atom by Thomson in 1904 \cite{Thomson04a}.
Since then, the ``Thomson problem'' has come to refer to finding
the energetically optimal
distribution of point charges constrained to the surface of a sphere.  This
idealised model has received considerable attention as an interesting exercise in
global optimisation \cite{Wales06b}.  Although our charged Lennard-Jones cluster
seems to share some common features with the Thomson problem, we note that the charged
particles would only be free to move continuously on the cluster surface when the
cluster has melted, since at low temperatures the charges would be constrained to
lie at the sites of the icosahedral structure.  However, at liquid-like temperatures
we have seen that the charges do not all reside at the surface of the sphere and
that entropy plays a large role in determining their distribution.  This means that
it is not easy to make contact with the Thomson problem in this context.
\par
Returning to the original motivation of a better understanding of electrospray
ionisation, it will now be essential to move to a dynamical treatment of cluster
instability and the mechanisms of fragmentation.  An unconfined cluster undergoes
evaporation even of neutral particles, which would have the effect of concentrating
charge within a shrinking cluster.  It will therefore be important to approach the
instability from below as well as looking at the highly-charged regime.  The
Lennard-Jones model should allow such processes to be investigated in detail
for a cluster bound by simple interactions.  We intend to develop this coarse-grained
approach in a number of directions, including inhomogeneous droplets in which a
solute has been introduced.

\section*{Acknowledgments}
MAM thanks EPSRC (U.K.) for financial support.  DAB acknowledges support
from G{\'e}nop{\^o}le-Evry for postdoctoral funding as well as
the supercomputing centre of Champagne-Ardenne (ROMEO) for computational resources and
technical support.
MPG is grateful to Churchill College, Cambridge (U.K.) and the French Foreign Office
for an Overseas Fellowship.  Travel associated with this collaboration was supported by the Alliance
Programme of the British Council and the French Minist{\`e}re des Affaires Etrang{\`e}res.
The authors thank Dr Florent Calvo, Prof.~Daan Frenkel and Dr Benjamin Rotenberg
for helpful discussions.

\section*{References}

\end{document}